\documentstyle[12pt]{article}

\fnsymbol{footnote}

\title {Relativistic quark model and meson Regge trajectories  }
\author{S.M.~Gerasyuta \footnote{Present address: Department of Physics,
 LTA, Institutski Per.5, St.Petersburg 194021, Russia }}
\date{}

\begin{document}
\maketitle
\centerline{Department of Theoretical Physics, St. Petersburg }
\centerline{State University, 198904, St.Petersburg, Russia}
\vspace{2cm}

\begin{abstract}

The D-and F-wave quark-quark amplitudes are
constructed in the framework of the dispersion
N/D-method. The mass values of meson multiplets
with $J^{PC}=1^{--},2^{--},2^{-+},3^{--}$ and
$J^{PC}=2^{++},3^{++},3^{+-},4^{++}$ are 
calculated.

   The Regge trajectories of mesonic resonances
with orbital numbers L=0,1,2,3 are obtained.

\end{abstract}

\newpage

{\large \bf I. INTRODUCTION}
\vspace{0.5cm}

    In soft processes, where small momentum transfers are essential, the
 perturbative QCD technique is not applicate. In this case the
way of phenomenology based on QCD seems to be
reasonable. One believes that such a phenomenology
should include quark model results as well. In the
framework of quark models there was obtained the important information on
light-quark mesons [1-9] and baryons [10-16].

    In the framework of the dispersion N/D-method
with help of the iteration bootstrap procedure the scattering amplitudes of
dressed quarks were constructed [17,18]. The mass values of the lowest
mesons($J^{PC}=0^{-+},1^{--},0^{++}$) and their
quark content are obtained. In the colour meson channel a bound state was found
which corresponds
to the constituent gluon with the mass $M_{G}=$0.67~GeV. The qq-amplitudes in
the colour state $\overline{3_{c}}$ have the diquark levels with $J^{P}=0^{+}$
and the masses $m_{ud}=$0.72~GeV and $m_{us}=m_{ds}=$0.86~GeV.
The interaction of dressed quarks appeared to be an effectively short-range
one. The calculated amplitudes satisfy the Okubo-Zweig-Iisuka rule.
The creation of mesons (pion included) is mainly
due to the gluon exchange. The model under consideration proceeds from the
assumption that the quark interaction forces are the two-component ones.
The first, short-range component
corresponds to the gluon exchange, the second, long-range component is due to
the confinement. When the low-lying mesons are
considered, the long-range component of the forces is neglected. But for the
excited mesons the long range forces are important. Namely, the confinement
of the $q\overline{q}$ pair with comparatively large energy is actually
realized as the production of the new $q\overline{q}$ pairs. This
means that in the transition $q\overline{q}\rightarrow q\overline{q}$
the forces appear which are connected with the contribution of box-diagrams
[19]. These box-diagrams can be important in the formation of hadron spectra.
We take into account the contribution of box-diagrams. It allows us to calculate
the mass spectrum of P-wave mesons in the relativistic quark model [20].

    In the present paper in the framework of relativistic quark model the mass
values of D- and F-wave mesons are calculated. The Regge trajectories for the
low-energy region, which describe the mesonic resonances with orbital numbers
L=0,1,2,3, are constructed.

    In section 2 the D- and F-wave quark-quark amplitudes in the framework of
dispersion N/D-method are constructed. The mass values of meson multiplets
with $J^{PC}=1^{--},2^{-+},2^{--},3^{--}$ and $J^{PC}=2^{++},3^{+-},3^{++},4^
{++}$ are calculated. The technique of the N/D-method procedure is presented
in the Appendix. In the Conclusion the status of the considered model is
discussed.

\vspace{1cm}
{\large \bf II. D- AND F-WAVE QUARK-QUARK AMPLITUDES}
\vspace{0.5cm}

    In the papers [17,18,20] we considered the scattering amplitudes of the
constituent quarks of three flavours (u,d,s). The poles of these amplitudes
determine the masses of the light mesons. The masses of the constituent quarks
u and d are of the order of 300-400~MeV, the strange quark is
100-150~MeV heavier. The constituent quark is a
colour triplet and quark amplitudes obey the global colour symmetry. The gluon
interaction is assumed to be short-range. The quark scattering
amplitudes for singlet colour states can be written as:

\begin{equation} A(t,z)= \sum_{i} G^{i}(t,z)(\overline{q}O^{i} q)
(\overline{q'}O^{i}q'),  \end{equation}
where $ O^{i}$ is a full set of matrices 4x4 (see
(A4)). z is cosine of the scattering angle in c.m.s.
    Using the general amplitude $ A(t,z)$ one can obtain the l-wave part
$ A_{l}(t)$, which determines the contributions of the l-wave quark-quark
amplitudes ((A1)-(A5)). Then we must expand the amplitude $ A_{l}(t) $ into
the eigenstates ((A6)-(A8)) and calculate the first approach amplitudes with
help of dispersion N/D-method ((A9)-(A14)) [17,18]. The poles of amplitudes
$ A_{l}(t)$ correspond to the value mass of l-wave meson multiplets.
The detailed exposition of the construction of D- and F-wave quark-quark
amplitudes in the Appendix is given.

  The N-function of the first approach for the calculation of low-lying meson
spectra (Table I) are used [17,18]. $ N_{i}(t)$-functions depend weakly on the
energy and therefore can be parametrized in our case:
$$ N^{S}(t)=-0.728-\frac{2.85}{t+2.92}, N^{A}(t)=-0.318-\frac{0.798}
{t+3.05},  $$
\begin{equation} N^{V}(t)=0.336+\frac{1.50}{t+2.60},
N^{P}(t)=0.690+\frac{2.86}{t+3.30},  \end{equation}
$$ N^{T}(t)\approx 0 $$

    These function was obtained by help of iteration bootstrap procedure with
the four-fermion interaction as an input [17,18]:

\begin{equation}   g_{V}(\overline{q}\gamma_{\mu}\lambda^{a}q)
(\overline{q}'\gamma_{\mu}\lambda^{a}q'),     \end{equation}
where $ \lambda^{a}$ are the Gell-Mann matrices. The point-like structure of
this interaction is motivated by the above-mentioned idea of the two
characteristic sizes in the hadron. On the other hand, the aplicability of
(eq.3) is verified by the success of the De Rujula-Georgi-Glashow quark
model [1], where only the short-range part of the Breit-Fermi potential
connected with the gluon exchange Fig.1(a) is responsible for the mass
splitting in hadron multiplets. But for the excited mesons the long-range
forces are important. The box-diagrams Fig.1(b) can be important in the
formation of hadron spectra.
We do not see any difficulties in taking into account the box-diagram with
help of the dispersion technigue. For the sake of simplicity one restrict
to the introduction of quark mass shift $\Delta_{l}$, which are defined by the
contribution of the nearest production thresholds of pair mesons $\pi \pi,
 \pi \eta, K \overline{K}, K \eta $ and so on.
We suggest that the parameter $ \Delta_{l}$ takes into account the confinement
potential effectively: $m^{*}=m+\Delta_{l}$, $m^{*}_{s}=m_{s}+\Delta_{l}$ and
changes the behaviour of pair quark amplitudes. It allows us to construct the
D- and F-wave mesons amplitudes and calculate the mass spectra by analogy with
the P-wave meson spectrum in the relativistic quark model (Table II) [20].
The calculated values of mass D- and F-wave mesons are shown in the
Tables III, IV respectively.
One can see that the obtained mass values of the D-wave mesons
$(J^{PC}=1^{--},2^{-+},2^{--},3^{--})$ are in good
agreement with experimental ones [21]. The absence of experimental data for
the F-wave mesons does not allow to verify the detailed coincidence.
The $\Delta_{l}$-parameters can be determined by
mean of fixing of masses $\rho_{3}$ (D-wave) and
$f_{4}$ (F-wave) mesons: $\Delta_{D}=$0.460~GeV, $\Delta_{F}=$0.640~GeV
respectively. The calculated mass values of D-and F-wave mesons are in good
agreement with the other model results [7-9].

    Using the obtained mass values of S- and P-wave meson multiplets [18,20]
and the results of the present paper, one can construct the meson Regge
trajectories $\alpha_{l}(t)=\alpha_{l}(0)+\alpha'_{l}t$ for the mesonic
resonances with orbital numbers L=0,1,2,3 (Table V). These results
are in good agreement with experimental data [21]
and other papers results [22-24]. The corresponding Regge trajectories
$(\rho,\omega; a_{0},f_{0}; a_{1},f_{1})$ have the exchange degeneracy,
excluding the $\alpha_{\pi}$ and $\alpha_{\eta}$ trajectories. The
$\alpha_{\pi}$-Regge trajectories differs from the straight only in the
low-energy range.

\vspace{1cm}
{\large \bf III. CONCLUSION}
\vspace{0.5cm}

    In the framework of the approach developed for the S- and P-wave mesonic
multiplets [18,20], the D- and F-wave mesons are calculated. In present
paper we suggest also, that the parameters $\Delta_{l}$ take into account the
contribution of box-diagram Fig.1(b). This is more important, that in the
framework of $1/N_{C}$-expansion [25,26] both diagrams Fig.1(a,b) have the equal
order $\sim 1/N_{C}$, where $N_{C}=N_{f}$ are the colour and flavour numbers
respectively.
   In the recent papers [27,28] the high energy asymptotic of multi-colour QCD
is considered, that allows to construct the Bete-Salpeter equation for the n
reggeized gluons. Then the author
obtained the Pomeron trajectory as the bound state of two reggeized gluons.
Therefore is important to obtained the meson Regge trajectories for the
low-energy region in the framework of the relativistic quark model, which is
based on the principles of multi-colour QCD.

\vspace{0.8cm}
\centerline{\large \bf ACKNOWLEDGMENTS}
\vspace{0.5cm}
   I am indebted to T.~Barnes, V.A.~Franke, L.N.~Lipatov, Yu.V.~Novozhilov
   for useful discussions.

\vspace{1cm}
\centerline{\large \bf APPENDIX}
\vspace{0.5cm}

   The calculation of the quark-quark amplitude for the singlet colour states
with the accounting of high excited states for the orbital numbers L=0,1,2,3
is performed using the dispersion N/D-method:

$$A(t,z)=\sum_{i}(N^{i}_{S,D}(t)+zN^{i}_{P,F}(t)+
z^{2}N^{i}_{D}(t)+z^{3}N^{i}_{F}(t))M^{i}_{t}=              $$
$$ =A_{S,D}(t)+A_{P,F}(t)+A_{D}(t)+A_{F}(t),  \eqno (A1)    $$
where $$N^{i}_{l}(t)=\int^{1}_{-1}\frac{dz}{2} P_{l}(z)
G^{i}(t,z), l=S,P,D,F   \eqno (A2) $$

The estimation of the contributions of high exitation give rise to the
renormalization of input parameters of S- and P-wave meson amplitudes, does
not change the mass spectra of these mesons.
    Here we introduce the matrix element $M^{i}_{t}$:
$$M^{i}_{t}=(\overline{q}O^{i}q)(\overline{q'}O^{i}q'), \eqno  (A3) $$
where $ O^{i}$ are the operators of different types of the four-fermion
interaction (i=S,V,T,A,P):

$$O^{i}=1,\gamma_{\mu},\frac{i}{\sqrt{2}}\sigma_{\mu\nu},
i\gamma_{\mu}\gamma_{5},\gamma_{5}  \eqno  (A4) $$
For the brevity the colour and flavour indices in the equation (A3) are
omitted. For the processes with quarks of different flavours the amplitude
should contain the sixth invariant,

$${k'}_{\mu}(\overline{q}\gamma_{5} {\gamma}_{\mu} q)(\overline{q'} \gamma_{5}
q')+k_{\mu}(\overline{q}\gamma_{5} q)(\overline{q'}\gamma_{5} \gamma_{\mu} q'),
\eqno (A5) $$
where $k_{\mu}$ and $k'_{\mu}$ are relative momenta of quarks from the
initial and final states. Our calculation shows that the contribution of the
sixth invariant into the quark-quark
amplitudes is small and can be neglected. Using the dispersion
N/D-method [17,18,20] one can expand the D- and F-wave amplitude in the
t-channel $q\overline{q}$ states in the following way:

$$A_{D}(t)=\sum_{J^{PC},i,\rho\rho',\sigma\sigma'}M^{i}_{t\rho\rho'\sigma
\sigma'}(J^{PC})N^{i}_{D}(t)n_{\rho}n_{\rho'}n_{\sigma}n_{\sigma'}
\eqno (A6)  $$
$$A_{F}(t)=\sum_{J^{PC},i,\rho \rho',\sigma \sigma',\lambda \lambda'}
M^{i}_{t\rho \rho' \sigma \sigma' \lambda \lambda'}(J^{PC})N^{i}_{F}(t)
n_{\rho} n_{\rho'} n_{\sigma} n_{\sigma'} n_{\lambda} n_{\lambda'},$$
where $ \vec n$ is the unity vector directed along
the relative momentum in an intermediate state.
    Futher we consider the construction of D-wave
amplitude in detail. The F-wave amplitude can be obtained analogously.
    Here the matrix element $M^{i}_{t\rho\rho'\sigma\sigma'}(J^{PC})$ is used:
$$ M^{i}_{t \rho \rho' \sigma \sigma'}(J^{PC})=(\overline{q}O^{i}q)
d^{i}_{\rho \rho' \sigma \sigma'}(J^{PC})(\overline{q'}O^{i}q')  \eqno (A7) $$
   The projectors $d^{i}_{\rho\rho'\sigma\sigma'}(J^{PC})$  for the D-wave
are defined as:

$$d^{S}_{\rho\rho'}(1^{--})=D_{\rho\rho'},
d^{P}_{\rho\rho'\sigma\sigma'}(2^{-+})=D_{\rho\rho'}D_{\sigma\sigma'} , $$
$$d^{A}_{\mu\mu'\rho\rho'\sigma\sigma'}(2^{-+})=\Pi_{\mu\mu'}D_{\rho\rho'}D_
{\sigma\sigma'}, $$
$$d^{V}_{\mu\mu'\rho\rho'\sigma\sigma'}(2^{--})=d^{V}_{\mu\mu'\rho\rho'}
(1^{++})D_{\sigma\sigma'}, $$
$$d_{\mu\mu'\rho\rho'\sigma\sigma'}(3^{--})=d^{V}_{\mu\mu'\rho
\rho'}(2^{++})D_{\sigma\sigma'}, \eqno (A8) $$
$$d^{V}_{\mu\mu'\rho\rho'}(1^{++})=\frac{1}{2}D_{\mu\mu'} D_{\rho\rho'}-\frac{1}
{2}D_{\mu\rho'} D_{\mu'\rho}, $$
$$d^{V}_{\mu \mu' \rho \rho'}(2^{++})=\frac{1}{2}D_{\mu\mu'}D_{\rho \rho'}-
\frac{1}{3}D_{\mu\rho} D_{\mu' \rho'}+\frac{1}{2}D_{\mu\rho'} D_{\mu' \rho} $$
Here one use the projectors of P-wave states:
$d^{V}_{\mu\mu'\rho\rho'}(1^{++})$ and $d^{V}_{\mu\mu'\rho\rho'}(2^{++})$,
$\Pi_{\mu\mu'}=P_{\mu}P_{\mu'}/P^{2}$, $D_{\mu\mu'}=\delta_{\mu\mu'}-\Pi_{\mu
\mu'}$, P is the total momentum of quark pair.

    The interaction amplitude can be written as N/D-relation.
The expression for the amplitude expanded into the eigenstates have the
following form:

$$ A^{i}_{D}(t,J^{PC})=N^{i}_{D}(t)\frac{{\overline{D}}^{i}_{D}
(t,J^{PC})}{detD_{D}(t,J^{PC})}, \eqno (A9) $$
where$\overline{D_{D}} $ is the co-factor matrix $D_{D}$.
To obtain unitarized amplitude the elastic-unitarity condition is used, which
provides us with an imaginary part of the D-wave amplitude of the loop
diagram:

$$16\pi\sum_{J^{PC}} Im D^{i}_{D}(t,J^{PC})M^{i}_{t\rho\rho'\sigma\sigma'}
(J^{PC})= $$
$$=-\frac{[(t-(m^{*}_{1}+m^{*}_{2})^{2})(t-(m^{*}_{1}-m^{*}_{2})^{2})]^{1/2}}
{t} \times $$
$$\times \int \frac{d\Omega}{4\pi} Tr(O^{i}(\widehat{p_{1}}+
m^{*}_{1})O^{i}
(-\widehat{p_{2}}+m^{*}_{2}))n_{\rho}n_{\rho'}n_{\sigma}n_{\sigma'}N^{i}_{D}(t)
M^{i}_{t},   \eqno (A10) $$
$m^{*}_{1} $ and $ m^{*}_{2} $ are the effective quark and antiquark masses
respectively. Using the equation (A10)
we can extract the eigenstates for the D-wave amplitude:
$$ ImD^{i}_{D}(t,J^{PC})=-\rho^{i}_{D}(t,J^{PC})N^{i}_{D}(t)
\eqno (A11) $$
There one introduce the two-particle phase space for the unequal quark masses:
$$\rho^{i}_{D}(t,J^{PC})=\left(\alpha^{i}_{D}(J^{PC})\frac{t}{(m^{*}_{1}+m^{*}_
{2})^{2}}+
\beta^{i}_{D}(J^{PC})+\frac{(m^{*}_{1}-m^{*}_{2})^{2}}{t}\delta^{i}_{D}(J^{PC})
\right)\times  $$
$$\times \frac{[(t-(m^{*}_{1}+m^{*}_{2})^{2})(t-(m^{*}_{1}-
m^{*}_{2})^{2})]^{1/2}}{t}  \eqno (A12) $$

The coefficients $ \alpha^{i}_{D}(J^{PC})$, $\beta^{i}_{D}(J^{PC}) $,
$\gamma^{i}_{D}(J^{PC}) $
are given in Table VI. Using the dispersion relation with the cut-off,we define
the Chew-Mandelstam function [29]:

$$D^{i}_{D}(t,J^{PC})=1+\frac{1}{\pi}\int^{\Lambda}_{(m^{*}_{1}+
m^{*}_{2})^{2}}dt'
\frac{ImD^{i}_{D}(t',J^{PC})}{t'-t}  \eqno (A13) $$
$ N_{J^{PC}}(t)$-function for the D-wave quark-quark amplitudes are defined
by (eq.(2)):

$$N_{1^{--}}(t)=N^{S}(t) $$
$$ N_{2^{--}}(t)=N^{V} (t) $$
$$ N_{3^{--}}(t)=N^{V} (t) $$
$$ N_{2^{-+}}(t)=N^{P}(t)+\frac{(m^{*}_{1}+m^{*}_{2})^{2}}{t}
N^{A}(t)       \eqno (A14) $$
By analogy with (A6)-(A14) one can cosider the F-wave amplitudes.

\newpage
\begin{center}
 TABLE I.
Masses of lowest meson multiplets (GeV)
\vspace{0.3cm}

\begin{tabular}{|c|l|c|l|c|l|} \hline

$J^{PC}=0^{-+}$& Masses &$J^{PC}=1^{--}$ & Masses &$J^{PC}=0^{++}$& Masses  \\
\hline
$ \pi $  & 0.14(0.14) & $\rho $  & 0.77(0.77) & $a_0$  & 0.78(0.98)  \\
$ \eta $ & 0.48(0.55) & $ \omega $ & 0.77(0.78) & $f_0$ & 0.87(0.98)  \\
$ K $    & 0.50(0.50) & $K^{*} $  & 0.89(0.89) & $K_{0}^{*}$ & 0.88(1.35)  \\
$ \eta' $ & 1.00(0.96) & $\Phi $  & 1.00(1.02) & $f_0$ & 1.16(1.30) \\  \hline
\end{tabular}
\end{center}

\noindent
Parameters of model: cut-off $ \Lambda$=17.3,gluon
constant $ g_{V}$=0.226, four-fermion interaction induced by instantons
constant $ g_{I}$=-0.081, quark masses $ m_{u,d}$=0.385~GeV, $m_{s}$=0.501~GeV,
parameter of quark mass shift $\Delta_{S}$=0. Experimental values are
given in parentheses [21]

\vspace{1cm}
\begin{center}
 TABLE II.
Masses of three P-wave meson multiplets (GeV)
\vspace{0.3cm}

\begin{tabular}{|c|l|c|l|c|l|} \hline

$J^{PC}=1^{++}$& Masses &$J^{PC}=1^{+-}$& Masses & $J^{PC}=2^{++}$& Masses  \\
\hline
$ a_1 $ & 1.273(1.260) & $b_1 $ & 1.203(1.235) & $a_2$ & 1.320(1.320)  \\
$ f_1 $ & 1.273(1.285) & $ h_1 $ & 1.203(1.170) & $f_2$ & 1.320(1.270)  \\
$ K_{1}^{*} $& 1.385(1.400) &$K_{1} $& 1.308(1.270) &$K_{2}$& 1.436(1.430)  \\
$ f_1 $ & 1.497(1.420) & $h_{1}' $ & 1.414( -- ) & $f_2$ & 1.552(1.525)  \\
\hline
\end{tabular}
\end{center}

\noindent
Parameters of model: cut-off $ \Lambda$=17.3,gluon
constant $ g_{V}$=0.226, quark masses $ m_{u,d}$=0.385~GeV, $m_{s}$=0.501~GeV,
parameter of quark mass shift $\Delta_{P}$=0.275~GeV. Experimental values are
given in parentheses [21]

\newpage

\begin{center}
 TABLE III.
Masses of D-wave meson multiplets (GeV)
\vspace{0.3cm}

\begin{tabular}{|c|l|c|l|} \hline

$J^{PC}=1^{--}$& Masses       & $J^{PC}=2^{-+}$ &   Masses \\  \hline
$ {\rho}_1 $   & 1.590(1.700) & $\pi_{2}$       & 1.620(1.670) \\
$ {\omega}_1 $ & 1.590(1.600) & $ \eta_{2}$     & 1.620( -- )  \\
$ K_{1}^{*} $  & 1.690(1.680) & $K_{2}$         & 1.720(1.770)  \\
$ {\Phi}_1 $   & 1.800(1.680) & $\eta_{2}$      & 1.840( -- )  \\  \hline
\hline
  $J^{PC}=2^{--}$ & Masses       & $J^{PC}=3^{--}$ & Masses  \\  \hline
   $ \rho_{2}$    & 1.670( -- )  & $\rho_{3}$      & 1.690(1.690)  \\
   $ \omega_{2}$  & 1.670( -- )  & $\omega_{3}$    & 1.690(1.670)  \\
   $ K_{2}^{*}$   & 1.780(1.820) & $K_{3}^{*}$     & 1.800(1.780)  \\
   $ \Phi_{2}$    & 1.900( -- )  & $\Phi_{3}$      & 1.920(1.850)  \\  \hline
\end{tabular}
\end{center}

\noindent
Parameters of model are analogous  Table II, except of the quark mass shift
$\Delta_{D}$=0.460~GeV.

\begin{center}
 TABLE IV.
Masses of F-wave meson multiplets (GeV)
\vspace{0.3cm}

\begin{tabular}{|c|l|c|l|} \hline

$J^{PC}=2^{++}$ & Masses       & $J^{PC}=3^{+-}$ & Masses  \\  \hline

$a_{2}$         & 1.930( --  ) & $ b_{3} $       & 1.950( --  )  \\
$f_{2}$         & 1.930(2.010) & $ h_{3} $       & 1.950( --  )  \\
$K_{2}^{*}$     & 2.030( --  ) & $ K_{3} $       & 2.060( --  )  \\
$f_{2}$         & 2.140( --  ) & $ h_{3} $       & 2.170( --  )  \\  \hline
\hline
$J^{PC}=3^{++}$ & Masses       & $J^{PC}=4^{++}$ & Masses  \\  \hline

$a_{3}$         & 1.960( --  ) & $ a_{4} $       & 2.050( --  )  \\
$f_{3}$         & 1.960( --  ) & $ f_{4} $       & 2.050(2.050)  \\
$K^{*}_{3}$     & 2.070( --  ) & $ K^{*}_{4}$    & 2.160(2.045)  \\
$f_{3}$         & 2.180( --  ) & $ f_{4}$        & 2.280( --  )  \\  \hline
\end{tabular}
\end{center}

\noindent
Parameters of model are analogous  Table II, except of the quark mass shift
$\Delta_{F}$=0.640~GeV.

\newpage
\begin{center}
 TABLE V.
Regge trajectories $q\overline{q}$-mesons
\vspace{0.3cm}

\begin{tabular}{|c|l|l|} \hline

Regge trajectories     & $\alpha (0) $ & $ \alpha' $  \\  \hline

$\alpha_{\rho,\omega}$   &   0.5(0.5)    & 0.9(0.9)  \\
$\alpha_{K^{*}} $        &   0.4(0.4)    & 0.8(0.8)  \\
$\alpha_{\varphi}$       &   0.2(0.1)    & 0.8(0.9)  \\
$\alpha_{\pi}$           &   0. (0.)     & 0.8(0.8)  \\
$\alpha_{K}$             &  -0.3(-0.3)   & 0.7(0.7)  \\
$\alpha_{\eta}$          &  -0.2(-0.2)   & 0.8(0.8)  \\
$\alpha_{\eta^{'}}$      &  -0.6( -- )   & 0.8( -- ) \\
$\alpha_{a_{0},f_{0}}$   &  -0.3(-0.5)   & 0.6(0.6)  \\
$\alpha_{K^{*}_{0}} $    &  -0.4( -- )   & 0.6( -- ) \\
$\alpha_{\tilde{f_{0}}}$ &  -0.6( -- )   & 0.6( -- ) \\
$\alpha_{a_{1},f_{1}}$   &  -0.4( -- )   & 0.8( --)  \\
$\alpha_{K_{1}}$         &  -0.5(-0.5)   & 0.7(0.7)  \\
$\alpha_{\tilde{f_{1}}}$ &  -0.6( -- )   & 0.7( -- ) \\  \hline
\end{tabular}
\end{center}

\noindent
Experimental values of the parameters Regge trajectories are given in
parentheses [21].

\begin{center}
 TABLE VI.
Coefficients of Chew-Mandelstam functions
\vspace{0.3cm}

\begin{tabular}{|l|c|c|c|} \hline

$J^{PC}$ &  $ \alpha_{D}$ & $ \beta_{D}$ & $ \gamma_{D}$  \\  \hline

$1^{--}$ &$-\frac{1}{2}$  &$\frac{1}{2}$                & 0      \\
$2^{-+}$ &$\frac{1}{2}$   &$-\frac{1}{2}e $             & 0      \\
$2^{--}$ &$\frac{4}{7}$   &$-\frac{1}{14}-\frac{3}{7}e$ &$\frac{1}{14}$   \\
$3^{--}$ &$\frac{2}{7}$   &$\frac{3}{14}-\frac{10}{7}e$ &$-\frac{3}{14}$  \\
\hline
\hline
$J^{PC}$ & $ \alpha_{F}$  & $\beta_{F}$                 & $ \gamma_{F}$  \\
\hline
$2^{++}$ &$-\frac{1}{2}$  &$\frac{1}{2}$                & 0  \\
$3^{+-}$ &$\frac{1}{2}$   &$-\frac{1}{2}e $             & 0  \\
$3^{++}$ &$\frac{11}{18}$ &$-\frac{1}{9}-\frac{7}{18}e$ &$\frac{1}{9}$  \\
$4^{++}$ &$\frac{5}{18}$  &$\frac{2}{9}-\frac{13}{18}e$ &$-\frac{2}{9}$  \\
\hline
\end{tabular}
\end{center}

\centerline{Here is $e={(m^{*}_{1}-m^{*}_{2})}^{2}/
{(m^{*}_{1}+m^{*}_{2})}^{2}$}

\newpage
{\large
\vspace*{2cm}

\unitlength 1mm
\thicklines
\begin{picture}(113.00,120.00)
\put(30.00,120.00){\vector(1,0){20.00}}
\put(50.00,120.00){\vector(1,0){40.00}}
\put(90.00,120.00){\line(1,0){21.00}}
\put(111.00,88.00){\vector(-1,0){21.00}}
\put(90.00,88.00){\vector(-1,0){40.00}}
\put(50.00,88.00){\line(-1,0){20.00}}
\put(70.00,92.00){\oval(6.00,8.00)[]}
\put(70.00,100.00){\oval(6.00,8.00)[]}
\put(70.00,108.00){\oval(6.00,8.00)[]}
\put(70.00,116.00){\oval(6.00,8.00)[]}
\put(75.00,104.00){\makebox(0,0)[lc]{$G$}}
\put(28.00,120.00){\makebox(0,0)[rc]{$q$}}
\put(28.00,88.00){\makebox(0,0)[rc]{$\bar q$}}
\put(113.00,120.00){\makebox(0,0)[lc]{$q$}}
\put(113.00,88.00){\makebox(0,0)[lc]{$\bar q$}}
\end{picture}     }

\vspace*{-8cm}
\noindent
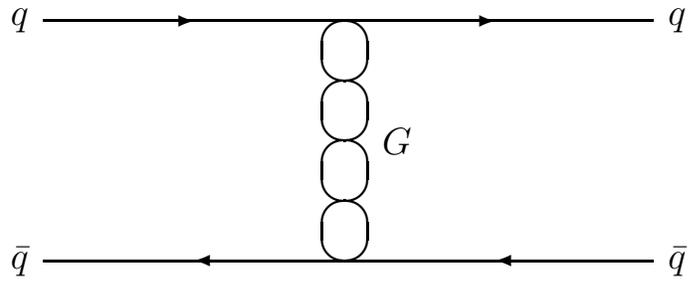
Fig.1 a) Diagram of gluonic exchange defines the short-range component of
quark interactions

\vspace*{5cm}

{\large
\unitlength 1.00mm
\thicklines
\begin{picture}(113.00,120.00)
\put(28.00,120.00){\makebox(0,0)[rc]{$q$}}
\put(28.00,88.00){\makebox(0,0)[rc]{$\bar q$}}
\put(113.00,120.00){\makebox(0,0)[lc]{$q$}}
\put(113.00,88.00){\makebox(0,0)[lc]{$\bar q$}}
\put(50.00,120.00){\vector(0,-1){16.00}}
\put(50.00,104.00){\line(0,-1){16.00}}
\put(90.00,88.00){\vector(0,1){16.00}}
\put(90.00,104.00){\line(0,1){16.00}}
\put(30.00,120.00){\vector(1,0){10.00}}
\put(40.00,120.00){\line(1,0){10.00}}
\put(30.00,88.00){\line(1,0){10.00}}
\put(50.00,88.00){\vector(-1,0){10.00}}
\put(90.00,120.00){\vector(1,0){10.00}}
\put(90.00,88.00){\line(1,0){10.00}}
\put(54.00,120.00){\oval(8.00,6.00)[]}
\put(62.00,120.00){\oval(8.00,6.00)[]}
\put(70.00,120.00){\oval(8.00,6.00)[]}
\put(78.00,120.00){\oval(8.00,6.00)[]}
\put(86.00,120.00){\oval(8.00,6.00)[]}
\put(54.00,88.00){\oval(8.00,6.00)[]}
\put(62.00,88.00){\oval(8.00,6.00)[]}
\put(70.00,88.00){\oval(8.00,6.00)[]}
\put(78.00,88.00){\oval(8.00,6.00)[]}
\put(86.00,88.00){\oval(8.00,6.00)[]}
\put(70.00,125.00){\makebox(0,0)[cb]{M}}
\put(70.00,93.00){\makebox(0,0)[cb]{M}}
\put(100.00,120.00){\line(1,0){11.00}}
\put(111.00,88.00){\vector(-1,0){11.00}}
\end{picture}     } 

\vspace*{-8cm}
\noindent
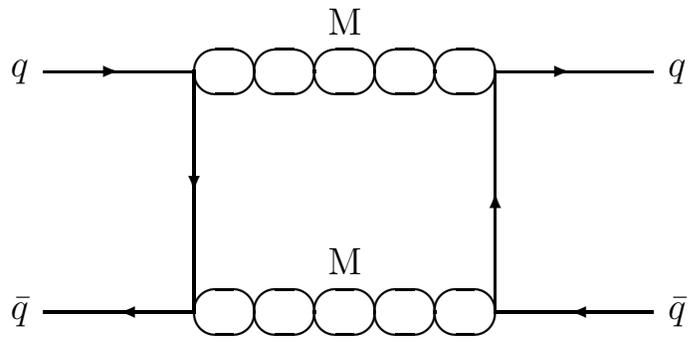
Fig.1 b) box-diagram of meson M takes into account the long-range interaction
component of the quark forces.

\end{document}